\documentclass[letterpaper, 10 pt, conference]{ieeeconf}  % Comment this line out
\IEEEoverridecommandlockouts                              % This command is only
\usepackage{orcidlink}
\usepackage[utf8]{inputenc}
\usepackage[T1]{fontenc}
\usepackage{graphicx}
\usepackage{cite}
\usepackage{hyperref}
\usepackage{amsfonts}
\usepackage[percent]{overpic}
\usepackage{subcaption}
\usepackage{amssymb}
\usepackage{amsmath}
\usepackage[capitalise]{cleveref}
\usepackage{mathtools}
\usepackage{breqn}
\usepackage{wrapfig}
\usepackage{xcolor}
\usepackage[percent]{overpic} % lets you place (x,y) in percent
\usepackage{float}
\usepackage{algorithm}
\usepackage{algpseudocode}
\usepackage{subcaption}
\usepackage{cite}
\usepackage{bm}
\usepackage{mathrsfs}
\DeclareMathAlphabet      {\mathbfit}{OML}{cmm}{b}{it}

\graphicspath{{Figures/}}

\title{\LARGE \bf
Graph-Based Characterization of Vision-Derived Dynamic Modes for Structural Damage Identification
}

\author{R K B M Rizmi, Khalid Mahmud Labib and Shabbir Ahmed$^{1} \orcidlink{0000-0001-8296-6025}$\\
Dynamical Signals and Systems Lab (DSSL)\\ Department of Mechanical Engineering\\
South Dakota State University, Brookings, SD % <-this % stops a space
\thanks{$^{1}$Corresponding author; Shabbir Ahmed, Assistant Professor, Mechanical Engineering, South Dakota State University, Brookings, SD, USA. Email {\tt\small shabbir.ahmed@sdstate.edu}}.%
}

\begin{document}

\maketitle
\thispagestyle{empty}
\pagestyle{empty}

%%%%%%%%%%%%%%%%%%%%%%%%%%%%%%%%%%%%%%%%%%%%%%%%%%%%%%%%%%%%%%%%%%%%%%%%%%%%%%%%
\begin{abstract}

%Identifying damage-sensitive features from non-contact measurements remains a key challenge in structural health monitoring. 
This study presents a graph theoretic framework for characterizing damage-induced changes in the dynamics of a vibrating cantilever beam from non-contact video measurements. Within this approach, dynamics of the beam under healthy and damaged conditions with crack depths of $5$, $10$, and $13$~mm are initially modeled using delay-embedded dynamic mode decomposition (DMD) from vision-based measurements. The resulting mode matrix from DMD is used to construct the adjacency matrix of a graph for this vibrating beam system, and its topological features such as graph connectivity, centrality, two-star statistics, and dyadic configurations are evaluated as damage-sensitive measures. Additionally, different DMD-based spectral and modal measures are evaluated to support the graph theoretic damage assessment. The results show a monotonic reduction in graph connectivity, nodal accessibility, and local organization of the graph with increasing crack depth. A parametric numerical study further validates the decreasing connectivity trend observed experimentally. The results demonstrate that the topological features of the graph derived from the DMD mode matrix provide an interpretable representation of damage-induced changes in the beam dynamics and offer a promising basis for vision-based structural damage identification.

\end{abstract}

%%%%%%%%%%%%%%%%%%%%%%%%%%%%%%%%%%%%%%%%%%%%%%%%%%%%%%%%%%%%%%%%%%%%%%%%%%%%%%%%
\section{Introduction} \label{sec:intro}

Identifying damage and damage-induced changes within a structure at multiple locations with conventional contact-based sensing is difficult because it requires a large number of sensors and associated instrumentation, increasing the overall cost and complexity of the monitoring system \cite{ahmed2022statistical,feng2017experimental,ahmed2025functional}. In this context, vision-based sensing may provide a non-contact alternative for damage detection in a structure \cite{feng2016vision}. Recent advances in video processing algorithms have enabled the extraction of full-field displacement or vibration responses from video data \cite{garrido2023damage,2026arXiv260502803R}. These measurements provide rich spatio-temporal information but generate high-dimensional data, requiring efficient data processing algorithms to extract the dominant dynamics or damage-sensitive features from data.

Previous studies have employed dynamic mode decomposition (DMD) for vibration characterization and damage identification of a vibrating beam from video data \cite{colombo2026damage,2026arXiv260502803R}. DMD represents the dynamics of a system through a linear mapping in a high-dimensional space whose spectral decomposition provides the corresponding eigenspectrum and mode matrix of the system \cite{schmid2010dynamic}. Most studies extract features from individual DMD modes for damage diagnosis, however, identification of individual modes that are sensitive to damage may be time-consuming for large data sets. Additionally, there may be multiple modes that are sensitive to damage and excluding some of them could result in the loss of potentially important information contained in the data. In this study, we treat the entire mode matrix as an evolving graph where the connection between its nodes and edges may change dynamically as damage alters the dynamics of the system.

 Dynamic graph models provide a framework for examining how network topology changes across different states of a dynamical system \cite{sayama2015introduction}. Previously, they were primarily applied in the social science field to analyze complex data, people, and their interactions \cite{sayama2015introduction,frank1986markov,freeman1977set}. In the present study, different structural conditions are represented as different graph states constructed from the corresponding DMD mode matrices, allowing changes in the identified dynamics to be characterized through variations in graph topology.
 
 %but its use has since expanded to physical systems, including fluid dynamics \cite{taira2016network,iacobello2019lagrangian}, thermal transport \cite{xiong2018influence}, power systems \cite{giacomarra2024generating}, and water-distribution networks \cite{giustolisi2019centrality}. However, to the best of our knowledge, network-theoretic approaches have not been used to characterize damage-induced changes in the dynamics of beam structures. In the present study, different structural conditions are represented as network states constructed from the corresponding DMD mode matrices, allowing changes in the identified dynamics to be characterized through variations in network topology.

%To support this graph-theoretic damage assessment, spectral measures from DMD algorithm, such as eigenspectrum eccentricity, Wasserstein, Chamfer, and Hausdorff distances, together with the similarity of matched DMD modes quantified using cosine similarity, are investigated. These quantities have been previously used in different fields, such as circularity of complex random variables\cite{ollila2008circularity}, neural-network training through comparison of Koopman spectra \cite{redman2024equivalent}, 3D shape completion from partial point clouds\cite{lin2024hyperbolic}, and vision-based structural vibration analysis using DMD and optDMD\cite{colombo2026damage}. However, to the best of our knowledge, these measures have not been used for damage diagnosis based on changes in the DMD eigenspectrum and modes.

To support the proposed graph-theoretic damage assessment, several spectral measures derived from the DMD eigenspectrum are investigated, including eigenspectrum eccentricity and the Wasserstein, Chamfer, and Hausdorff distances, together with the cosine similarity between matched DMD modes. These measures have previously been employed across a range of applications such as the characterization of circularity in complex random variables \cite{ollila2008circularity}, the identification of equivalent training dynamics in deep neural networks through comparisons of Koopman spectra \cite{redman2024equivalent}, the quantification of geometric dissimilarity between point clouds \cite{lin2024hyperbolic}, and vibration-based structural analysis using DMD and optDMD \cite{colombo2026damage}. In this study, these spectral and modal measures are employed as complementary indicators for damage assessment using vibration data extracted from beam videos.

The remainder of this paper is organized as follows: Section~\ref{theory} presents the theoretical background of dynamic mode decomposition and graph-theoretic measures used in this study, Section~\ref{sec:methods} describes the experimental methodology and numerical simulation of a beam under specific damage scenarios, and Section~\ref{sec:results} presents the obtained results and associated discussions. Section~\ref{sec:conclusion} discusses the conclusions.

\section{Theoretical Background}
\label{theory}

\subsection{DMD-Based Prediction}
In this study, delay-embedded dynamic mode decomposition (DMD) is applied to extract the beam dynamics from the video data. The beam centerline is tracked at 120 spatial locations by computing the midpoint between the detected top and bottom edges of the beam in each video frame. The transverse displacement is then obtained from the change in these midpoint positions relative to the equilibrium position. The measured displacement data are arranged into the matrix,
\begin{equation}
\resizebox{0.91\columnwidth}{!}{$
\mathbf{Y}
=
\begin{bmatrix}
y_{1}(t_{1}) & \cdots & y_{1}(t_{N}) \\
\vdots & \ddots & \vdots \\
y_{120}(t_{1}) & \cdots & y_{120}(t_{N})
\end{bmatrix}
=
\begin{bmatrix}
\mathbf{y}(t_1) & \cdots & \mathbf{y}(t_N)
\end{bmatrix}
\in\mathbb{R}^{120\times N}
$}
\label{eq:displacement_matrix}
\end{equation}
where $y_j(t_k)$ denotes the transverse displacement measured in pixels at
the $j$th spatial location ($j=1,\ldots,120$) at discrete time instant
$t_k$, $N$ is the total number of time steps, and
$\mathbf{y}(t_k)$ contains displacement data at all spatial location at time $t_k$.
The displacement snapshots are then arranged in a delay-embedded
representation as described in our previous work \cite{2026arXiv260502803R}. For an embedding
dimension $d$, the snapshot matrices are constructed as
\begin{equation}
\resizebox{0.91\columnwidth}{!}{$
\mathbf{X}
=
\begin{bmatrix}
\mathbf{y}(t_1) & \cdots & \mathbf{y}(t_{N-d}) \\
\vdots & \ddots & \vdots \\
\mathbf{y}(t_d) & \cdots & \mathbf{y}(t_{N-1})
\end{bmatrix},
 \mathbf{X}'
=
\begin{bmatrix}
\mathbf{y}(t_2) & \cdots & \mathbf{y}(t_{N-d+1}) \\
\vdots & \ddots & \vdots \\
\mathbf{y}(t_{d+1}) & \cdots & \mathbf{y}(t_N)
\end{bmatrix}
$}
\label{eq:delay_snapshot_matrices}
\end{equation}
where $\mathbf{X},\mathbf{X}'\in\mathbb{R}^{120d\times(N-d)}$. The DMD algorithm seeks a best-fit linear operator $\mathbf{A}$ that establishes a linear dynamical system to advance snapshot measurements forward in time and relates these snapshot matrices as
\begin{equation}
\mathbf{X}'\approx\mathbf{A}\mathbf{X}.
\label{eq:dmd_mapping}
\end{equation}
where $\mathbf{A} \in \mathbb{R}^{120d \times 120d}$. A truncated SVD of the snapshot matrix, $\mathbf{X} \approx \mathbf{U}_r \boldsymbol{\Sigma}_r \mathbf{V}_r^*$, retains the leading $r$ singular triplets, with $\boldsymbol{\Sigma}_r = \mathrm{diag}(\sigma_1, \ldots, \sigma_r)$. Projecting the full-order operator onto this subspace yields
\begin{equation}
\tilde{\mathbf{A}}
=
\mathbf{U}_r^* \mathbf{A} \mathbf{U}_r
=
\mathbf{U}_r^* \mathbf{X}' \mathbf{V}_r \boldsymbol{\Sigma}_r^{-1}
\in \mathbb{C}^{r \times r},
\end{equation}
whose eigendecomposition $\tilde{\mathbf{A}} \mathbf{W} = \mathbf{W} \boldsymbol{\Lambda}$ gives the DMD eigenvalues $\boldsymbol{\Lambda} = \mathrm{diag}(\lambda_1, \ldots, \lambda_r)$ and eigenvectors $\mathbf{W}$. The exact DMD modes follow as $\boldsymbol{\Phi} = \mathbf{X}' \mathbf{V}_r \boldsymbol{\Sigma}_r^{-1} \mathbf{W}$~\cite{tu2014dmd}. The modes $\boldsymbol{\Phi}$ encode the spatial structure of the identified dynamics and therefore reflect changes in the system dynamics, such as crack-induced local stiffness reduction~\cite{sinha2002simplified}. Because delay embedding replicates the original observables~\cite{fujii2019spectral}, the mode matrix contains $d$ redundant blocks of size $120 \times r$. Averaging across these blocks yields a reduced mode matrix $\bar{\boldsymbol{\Phi}} \in \mathbb{C}^{120 \times r}$, which is then used as the adjacency matrix for the network-theoretic measures.

\begin{figure}[]
%\hspace*{-0.5cm}
\includegraphics[width=1\linewidth]{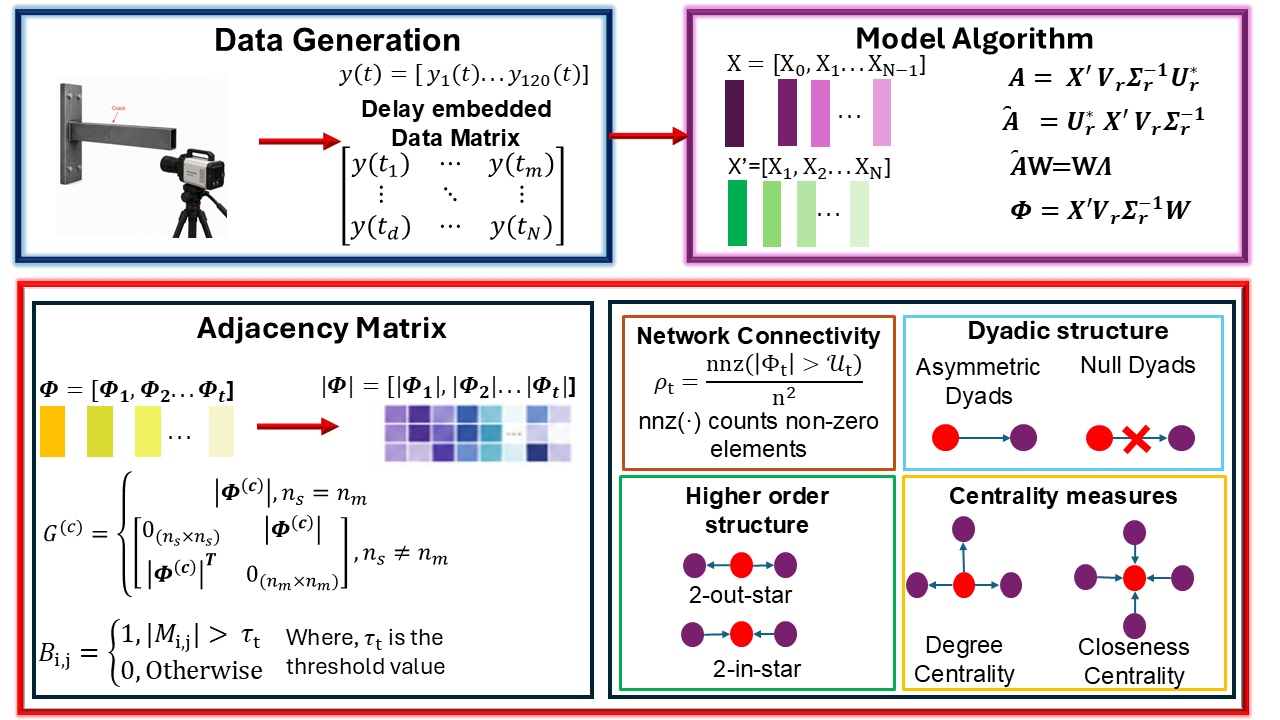}
\caption{Schematic representation of the proposed graph-theoretic framework for structural damage diagnosis from video data.}
\label{fig:Schematic_beam}
\vspace{-4mm}
\end{figure}

\subsection{Graph-Theoretic Measures} 
\label{sec:graph_measures}

In the proposed framework, the DMD mode matrix $\boldsymbol{\Phi}$ at each structural condition $c$ (healthy, 5 mm, 10 mm, and 13 mm crack depths) is used to construct the graph $\mathcal{G}^{(c)}$ as follows \cite{Labib2026Graph}:
\begin{equation}
\mathcal{G}^{(c)}
=
\left(
V,
E^{(c)},
\mathbf{G}^{(c)}
\right),
\end{equation}
where $V$ denotes the fixed set of nodes, $E^{(c)}$ denotes the condition-dependent set of edges, and $\mathbf{G}^{(c)}$ denotes the weighted adjacency matrix representing the configuration of the dynamic graph $\mathcal{G}^{(c)}$. The edge weights in $\mathbf{G}^{(c)}$ are obtained from the DMD mode matrix $\boldsymbol{\Phi}^{(c)}$. When the number of rows ($n_s$, spatial locations) and the number of columns ($n_m$, number of modes) of the $\boldsymbol{\Phi}^{(c)}$ are equal, the weighted adjacency matrix $\mathbf{G}^{(c)}$ is defined by its magnitude directly as

\begin{equation}
\mathbf{G}^{(c)}
=
\left|
\boldsymbol{\Phi}^{(c)}
\right|,
\qquad n_s=n_m,
\end{equation}
When $n_s\neq n_m$, the DMD mode matrix $\boldsymbol{\Phi}^{(c)}$ is rectangular and the weighted adjacency matrix $\mathbf{G}^{(c)}$ is constructed as follows \cite{dhillon2001coclustering}: 

\begin{equation}
\mathbf{G}^{(c)}
=
\begin{bmatrix}
\mathbf{0}_{n_s\times n_s} &
\left|\boldsymbol{\Phi}^{(c)}\right|\\
\left|\boldsymbol{\Phi}^{(c)}\right|^{T} &
\mathbf{0}_{n_m\times n_m}
\end{bmatrix},
\qquad n_s\neq n_m.
\end{equation}

Since the rows and columns of $\boldsymbol{\Phi}^{(c)}$ correspond to spatial locations and DMD modes, respectively, the resulting graph is naturally bipartite in both cases. When $n_s=n_m$, the square matrix $\left|\boldsymbol{\Phi}^{(c)}\right|$ is used directly as the weighted biadjacency representation. When $n_s\neq n_m$, the rectangular biadjacency matrix is embedded into a symmetric block adjacency matrix. For the subsequent graph theoretic analysis, the weighted adjacency matrix $\mathbf{G}^{(c)}$ of the graph $\mathcal{G}^{(c)}$ is converted into the binary adjacency matrix $\mathbf{B}^{(c)}$ by applying a threshold $\tau$. The threshold determines which weighted connections in $\mathbf{G}^{(c)}$ are retained in $\mathbf{B}^{(c)}$. With connections exceeding $\tau$ is assigned a value of one, and the remaining connections are assigned a value of zero. A threshold sensitivity analysis is used to determine a suitable value of $\tau$ for which the resulting network statistics maintain consistent trends across the structural conditions. The binary adjacency matrix $\mathbf{B}^{(c)}$ can be defined as follows:
\begin{equation}
B_{ij}^{(c)}
=
\begin{cases}
1, & G_{ij}^{(c)}>\tau,\\
0, & G_{ij}^{(c)}\leq\tau.
\end{cases}
\label{eq:graph_threshold}
\end{equation}
This binary adjacency matrix $\mathbf{B}^{(c)}$ is then used for the subsequent graph theoretic analysis, as illustrated in Fig.~\ref{fig:Schematic_beam}.

\subsubsection*{Graph Connectivity} The total number of active edges in the network or graph at structural condition $c$ is determined from the binary adjacency matrix $\mathbf{B}^{(c)}$ as $N^{(c)}$,
\begin{equation}
N^{(c)}
=
\sum_{i=1}^{n}\sum_{j=1}^{n} B_{ij}^{(c)},
\end{equation}
where $N^{(c)}$ denotes the total number of retained connections in the graph, corresponding to the number of nonzero entries in $\mathbf{B}^{(c)}$. A larger value of $N^{(c)}$ indicates that more connections are retained, whereas a smaller value indicates a reduction in active connections. To enable comparison across structural conditions, the graph connectivity $\rho^{(c)}$ is defined as the fraction of retained connections relative to the total number of possible connections,
\begin{equation}
\rho^{(c)}
=
\frac{N^{(c)}}{n^2}.
\label{simulation_connectivity}
\end{equation}
The network connectivity $\rho^{(c)}$ therefore provides a normalized, dimensionless measure of how densely the nodes in the network $\mathcal{G}^{(c)}$ are connected.

\subsubsection*{Dyadic and Higher-Order Structure}

Local graph structure is first characterized through dyadic configurations \cite{wasserman1996logit,frank1986markov}. A dyadic representation considers the relationship between each pair of nodes independently of the remaining node pairs in the network. For a pair of nodes $i$ and $j$, the corresponding dyad is defined as
\begin{equation}
D_{ij}
=
\left(B_{ij},B_{ji}\right),
\end{equation}
where $B_{ij}$ and $B_{ji}$ denote the directed edge weights from node $i$ to node $j$ and from node $j$ to node $i$, respectively. A dyad may be null, $D_{ij}=(0,0)$, asymmetric, $D_{ij}=(1,0)$ or $(0,1)$, or mutual, $D_{ij}=(1,1)$. 

Two-star configurations characterize connection patterns that share a common node in the network $\mathcal{G}^{(c)}$ at each structural condition $c$~\cite{wasserman1996logit,frank1986markov}. For node $i$, the out-degree and in-degree are
\begin{equation}
k_i^{\mathrm{out},(c)} = \sum_j B_{ij}^{(c)}, \qquad
k_i^{\mathrm{in},(c)}  = \sum_j B_{ji}^{(c)},
\end{equation}
from which the 2-out-star and 2-in-star statistics of $\mathcal{G}^{(c)}$ follow as
\begin{equation}
S_{\mathrm{out}}^{(c)} = \sum_i \binom{k_i^{\mathrm{out},(c)}}{2}, \qquad
S_{\mathrm{in}}^{(c)}  = \sum_i \binom{k_i^{\mathrm{in},(c)}}{2}.
\end{equation}
$S_{\mathrm{out}}^{(c)}$ counts pairs of outgoing edges sharing a common source node, while $S_{\mathrm{in}}^{(c)}$ counts pairs of incoming edges sharing a common target node.

% Two-star configurations are used to characterize connection patterns that share a common node in the network ${G}^{(c)}$ at each structural condition $c$ \cite{wasserman1996logit,frank1986markov}. For node $i$, the out-degree $k_i^{\mathrm{out},(c)}$ and in-degree $k_i^{\mathrm{in},(c)}$ are defined as
% \begin{equation}
% k_i^{\mathrm{out},(c)}
% =
% \sum_j B_{ij}^{(c)},
% \qquad
% k_i^{\mathrm{in},(c)}
% =
% \sum_j B_{ji}^{(c)}.
% \end{equation}
% Using these degree measures, the 2-out-star statistic $S_{\mathrm{out}}^{(c)}$ and 2-in-star statistic $S_{\mathrm{in}}^{(c)}$ of the network $\mathcal{G}^{(c)}$ are defined as
% \begin{equation}
% \begin{aligned}
% S_{\mathrm{out}}^{(c)}
% &=
% \sum_i
% \frac{
% k_i^{\mathrm{out},(c)}
% \left(k_i^{\mathrm{out},(c)}-1\right)
% }{2},
% \\[2mm]
% S_{\mathrm{in}}^{(c)}
% &=
% \sum_i
% \frac{
% k_i^{\mathrm{in},(c)}
% \left(k_i^{\mathrm{in},(c)}-1\right)
% }{2}.
% \end{aligned}
% \end{equation}
% The 2-out-star statistic $S_{\mathrm{out}}^{(c)}$ quantifies pairs of outgoing connections sharing the same node, whereas the 2-in-star statistic $S_{\mathrm{in}}^{(c)}$ quantifies pairs of incoming connections sharing the same node.

\subsubsection*{Centrality Measures}
To complement the dyadic and two-star statistics, node-level structural prominence is quantified using degree centrality $C_i^{D}$ and closeness centrality $C_i^{C}$. Together, these measures identify nodes that either interact with many others directly or occupy positions from which the rest of the network is readily accessible; both aspects are relevant when local damage alters the effective coupling between degrees of freedom. For brevity, the structural-condition index $(c)$ is suppressed throughout this subsection.

Degree centrality corresponds to the out-degree of node $i$ in the directed network, i.e., the number of outgoing connections originating from that node~\cite{sayama2015introduction}. It is defined as
\begin{equation}
C_i^{D} = k_i^{\mathrm{out}} = \sum_j B_{ij}, \qquad
\bar{C}^{D} = \frac{1}{n_x}\sum_i C_i^{D},
\end{equation}
where $\bar{C}^{D}$ is the network-averaged degree centrality and $n_x$ denotes the number of nodes. A node with high $C_i^{D}$ exerts direct influence on many other nodes, so a shift in $\bar{C}^{D}$ across conditions reflects a global change in the density of direct interactions.

Closeness centrality provides a complementary, path-based view by measuring how efficiently a node can reach the rest of the network through directed shortest paths. Larger values indicate shorter average path lengths to reachable nodes and, consequently, greater accessibility within the network. It is defined as
\begin{equation}
C_i^{C} = \frac{a_i^{\,2}}{(n_x-1)\displaystyle\sum_{j \in \mathcal{R}_i} d_{ij}}, \qquad
\bar{C}^{C} = \frac{1}{n_x}\sum_i C_i^{C},
\end{equation}
where $\mathcal{R}_i$ is the set of nodes reachable from node $i$, $a_i = |\mathcal{R}_i|$ is the number of nodes in this set, and $d_{ij}$ is the directed geodesic distance from $i$ to $j$. The factor $a_i^{2}$ in the numerator penalizes nodes that reach only a small fraction of the network, so $C_i^{C}$ jointly reflects both the reach and the proximity of node $i$. The network-averaged quantities $\bar{C}^{D}$ and $\bar{C}^{C}$ subsequently serve as scalar descriptors of each structural condition.

% \subsubsection*{Centrality Measures}

% To complement the dyadic and two-star statistics, node-level structural prominence is quantified using degree ($C_i^{D}$), and closeness centrality measures ($\bar{C}^{D}$). Degree centrality ($C_i^{D}$) represents the number of direct outgoing connections associated with node $i$ in the directed network \cite{sayama2015introduction}. It is defined as
% \begin{equation}
% C_i^{D}
% =
% k_i^{\mathrm{out}}
% =
% \sum_j B_{ij},
% \qquad
% \bar{C}^{D}
% =
% \frac{1}{n_x}\sum_i C_i^{D}.
% \end{equation}

% Closeness centrality ($\bar{C}^{D}$), provides a complementary measure based on the directed shortest-path distances between nodes. A larger value indicates shorter path distances to reachable nodes and, consequently, greater accessibility within the network. It is computed as
% \begin{equation}
% C_i^{C}
% =
% \frac{a_i^2}
% {(n_x-1)\displaystyle\sum_{j\in\mathcal{R}_i}d_{ij}},
% \qquad
% \bar{C}^{C}
% =
% \frac{1}{n_x}\sum_i C_i^{C},
% \end{equation}
% where $\mathcal{R}_i$ denotes the set of nodes reachable from node $i$, $a_i$ denotes the number of nodes in this set, and $d_{ij}$ represents the directed geodesic distance from node $i$ to node $j$.

\subsection{DMD-Based Measures for Damage Detection}
\label{fig:schematic_graph}

Five damage-sensitive parameters are derived from the DMD eigenspectrum and modes. Eigenspectrum eccentricity, Wasserstein distance, Chamfer distance, and Hausdorff distance are used as spectral indicators, while cosine similarity is used as a modal indicator. Following eigendecomposition of the reduced DMD operator, the discrete-time DMD eigenvalues are mapped to continuous-time poles according to
\begin{equation}
\omega_k
=
\frac{\log(\lambda_k)}{\Delta t}
=
\sigma_k+\mathrm{i}\Omega_k,
\label{eq:continuous_dmd_poles}
\end{equation}
where $\sigma_k$ denotes the growth or decay rate and $\Omega_k/(2\pi)$ denotes the corresponding oscillation frequency \cite{taira2020modal}. To place the real and imaginary components on comparable scales, the continuous-time poles are normalized using statistics obtained from the
healthy condition:
\begin{equation}
\mathbf{z}^{(c)}_i
=
\begin{bmatrix}
\operatorname{Re}\!\left(\omega_i^{(c)}\right)/s_{\sigma}\\
\operatorname{Im}\!\left(\omega_i^{(c)}\right)/s_{\Omega}
\end{bmatrix}
\in\mathbb{R}^{2},
\qquad i=1,\ldots,n_m,
\label{eq:scaled_poles}
\end{equation} where $n_m$ denotes the number of retained DMD modes, and $s_{\sigma}$ and $s_{\Omega}$ are the standard deviations of the real and imaginary components of the healthy-case poles, respectively. These scaling factors are held fixed for all structural conditions. The resulting scaled eigenspectrum at condition $c$ is represented by the matrix
\begin{equation}
\mathbf{Z}^{(c)}
=
\begin{bmatrix}
\mathbf{z}^{(c)}_1 &
\mathbf{z}^{(c)}_2 &
\cdots &
\mathbf{z}^{(c)}_{n_m}
\end{bmatrix}
\in\mathbb{R}^{2\times n_m}.
\end{equation}

\subsubsection*{Eigenspectrum Eccentricity}

Let $\gamma_1^{(c)}\geq\gamma_2^{(c)}$ denote the eigenvalues of the covariance matrix associated with the scaled spectral matrix $\mathbf{Z}^{(c)}$. The eigenspectrum anisotropy measure is defined as
\begin{equation}
e_c
=
\sqrt{
\frac{
\gamma_1^{(c)}-\gamma_2^{(c)}
}{
\gamma_1^{(c)}+\gamma_2^{(c)}
}
},
\label{eq:eccentricity}
\end{equation}
where larger values of $e_c$ indicate a more elongated distribution of
the retained DMD poles in the scaled spectral plane.

\subsubsection*{Spectral Distances}

Differences between the eigenspectrum at condition $c$ and the healthy reference are quantified using Wasserstein, Chamfer, and Hausdorff distances. For $n_m$ equally weighted spectral points, the second-order Wasserstein distance is defined as
\begin{equation}
W_2
\left(
\mathbf{Z}^{(0)},
\mathbf{Z}^{(c)}
\right)
=
\left[
\frac{1}{n_m}
\min_{\pi}
\sum_{i=1}^{n_m}
\left\|
\mathbf{z}^{(0)}_i
-
\mathbf{z}^{(c)}_{\pi(i)}
\right\|_2^2
\right]^{1/2},
\label{eq:wasserstein}
\end{equation} where $\pi$ denotes the optimal one-to-one assignment between the two
spectral point sets. The Chamfer distance measures the average nearest-neighbor discrepancy
between the two spectral point sets and is defined as
\begin{equation}
D_{\mathrm{C}}
\left(
\mathbf{Z}^{(0)},
\mathbf{Z}^{(c)}
\right)
=
\frac{1}{n_m}
\sum_{i=1}^{n_m}\min_j d_{ij}
+
\frac{1}{n_m}
\sum_{j=1}^{n_m}\min_i d_{ij},
\label{eq:chamfer}
\end{equation}
where
\begin{equation}
d_{ij}
=
\left\|
\mathbf{z}^{(0)}_i
-
\mathbf{z}^{(c)}_j
\right\|_2.
\end{equation}

The Hausdorff distance captures the largest nearest-neighbor discrepancy
between the two sets and is defined as
\begin{equation}
D_{\mathrm{H}}
\left(
\mathbf{Z}^{(0)},
\mathbf{Z}^{(c)}
\right)
=
\max
\left\{
\max_i\min_j d_{ij},
\max_j\min_i d_{ij}
\right\}.
\label{eq:hausdorff}
\end{equation} These three measures characterize complementary aspects of spectral
change. The Wasserstein distance uses an optimal one-to-one assignment between all retained poles, the Chamfer distance measures the average nearest-neighbor discrepancy, and the Hausdorff distance is governed by the largest nearest-neighbor discrepancy.

\subsubsection*{Matched-Mode Cosine Similarity}

To track modal changes across conditions, DMD modes at each structural condition $c$ are matched to the healthy reference through a joint criterion that combines spectral proximity and mode-shape similarity. The matching cost between the $i$th healthy mode and the $j$th mode at condition $c$ is
\begin{equation}
J_{ij}^{(c)} = \tfrac{1}{2}\left\|\mathbf{z}^{(0)}_i - \mathbf{z}^{(c)}_j\right\|_2^2 + \tfrac{1}{2}\left(1 - \mathrm{MAC}_{ij}^{(c)}\right),
\label{eq:joint_cost}
\end{equation}
where $\mathbf{z}$ denotes the DMD eigenvalue coordinates and $\mathrm{MAC}_{ij}^{(c)}$ is the modal assurance criterion~\cite{ardila2024automated}. Minimising $J_{ij}^{(c)}$ over $j$ yields the optimal assignment $\pi_c(i)$, and the average matched-mode cosine similarity follows as
\begin{equation}
S_{\Phi}^{(c)} = \frac{1}{n_m}\sum_{i=1}^{n_m}
\frac{\left|\left(\boldsymbol{\phi}^{(0)}_i\right)^H \boldsymbol{\phi}^{(c)}_{\pi_c(i)}\right|}
{\left\|\boldsymbol{\phi}^{(0)}_i\right\|_2 \left\|\boldsymbol{\phi}^{(c)}_{\pi_c(i)}\right\|_2},
\label{eq:cosine_similarity}
\end{equation}
where $S_{\Phi}^{(c)} \to 1$ indicates preservation of the healthy modal structure and lower values reflect increased modal change.

% \subsubsection*{Matched-Mode Cosine Similarity}

% The matching considers both spectral proximity and mode-shape similarity, where spectral proximity is evaluated from the distance between the corresponding spectral points and mode-shape similarity is quantified using the Modal Assurance Criterion (MAC) \cite{ardila2024automated}. The combined matching cost $J_{ij}$ between the $i$-th healthy mode and the $j$-th mode at structural condition $c$ is defined as
% \begin{equation}
% J_{ij}
% =
% 0.5\left\|\mathbf{z}^{(0)}_i-\mathbf{z}^{(c)}_j\right\|_2^2
% +
% 0.5\left(1-\mathrm{MAC}_{ij}\right).
% \label{eq:joint_cost}
% \end{equation} After optimal matching, let $\pi(i)$ denote the mode at structural condition $c$ matched to the $i$-th healthy mode. The cosine similarity $S_{\Phi,c}$ is then defined as the average similarity of the eight matched mode pairs,
% \begin{equation}
% S_{\Phi,c}
% =
% \frac{1}{8}
% \sum_{i=1}^{8}
% \frac{
% \left|
% \left(\boldsymbol{\phi}^{(0)}_i\right)^H
% \boldsymbol{\phi}^{(c)}_{\pi(i)}
% \right|
% }{
% \left\|
% \boldsymbol{\phi}^{(0)}_i
% \right\|_2
% \left\|
% \boldsymbol{\phi}^{(c)}_{\pi(i)}
% \right\|_2
% }.
% \label{eq:cosine_similarity}
% \end{equation}
% Values near unity indicate greater similarity to the healthy modes, whereas lower values indicate greater modal change.

\section{Data and Methodology}
\label{sec:methods}

To investigate the graph theoretic approach for damage diagnosis, a twofold scheme consisting of numerical and experimental analyses was employed. First, an in-house finite-element simulation of the corresponding $0.8$~m cantilever beam was performed for crack depths of $[0,2,\ldots,14]$~mm. The displacement response was obtained at 120 spatial locations and analyzed using delay-embedded DMD with $d=2$ and $r=29$. For the experimental analysis, videos were obtained from the open dataset of Garrido et al.~\cite{Garrido2024}. A polypropylene cantilever beam with a $0.8$~m free span and $25.4\times25.4$~mm cross-section was recorded under four structural conditions, consisting of the healthy case and crack depths of $5$, $10$, and $13$~mm. The analyzed videos had a frame rate of 60~fps. The visible beam span was sampled at 120 locations, and transverse displacement was obtained from the midpoint between the detected upper and lower beam edges. Delay-embedded DMD was applied with $d=60$ and $r=120$. The first 70\% of each response was used for model identification and damage-sensitive parameter calculation, while the remaining 30\% was used for prediction assessment.
% Because the experimental and numerical cases use different DMD and network configurations, only the damage-dependent connectivity trends are compared.

\section{Results and Discussion}
\label{sec:results}

%\subsection{Displacement Response and DMD Reconstruction}

\begin{figure}[t]
    \centering
 \vspace{5mm}
    \begin{overpic}[width=\columnwidth]
        {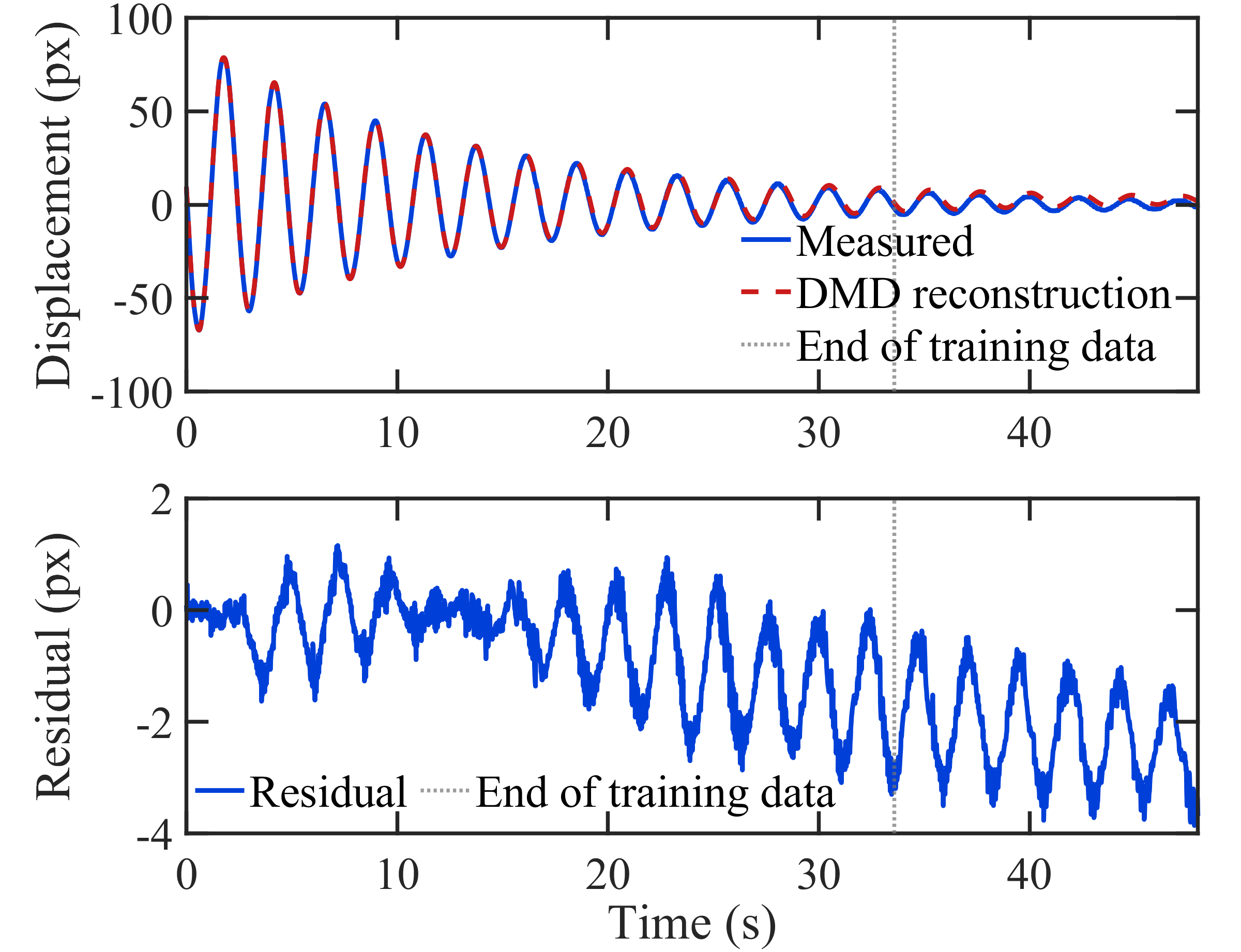}

        % Floating subfigure labels inside the image
        \put(89,71){\normalsize\textbf{(a)}}
        \put(89,32){\normalsize\textbf{(b)}}

    \end{overpic}

    \caption{Demonstration of DMD reconstruction performance: (a) measured tip displacement and the corresponding DMD reconstruction; (b) reconstruction residual. }

    \label{fig:dmd_reconstruction}
    \vspace{-5mm}
\end{figure}
First, the displacement responses extracted from the video data were modeled using delay-embedded DMD. Fig.~\ref{fig:dmd_reconstruction}(a) compares the measured tip displacement, shown in blue, with the corresponding DMD reconstruction, shown in red, while the dotted line indicates the end of the data used for model training. Fig.~\ref{fig:dmd_reconstruction}(b) presents the associated residual. The close agreement between the measured and reconstructed responses indicates that DMD captures the dominant beam dynamics well. We next investigated how the graph theoretic measures obtained from the DMD mode matrix evolve across the different structural conditions.

%\subsection{Graph Theoretic Measures}

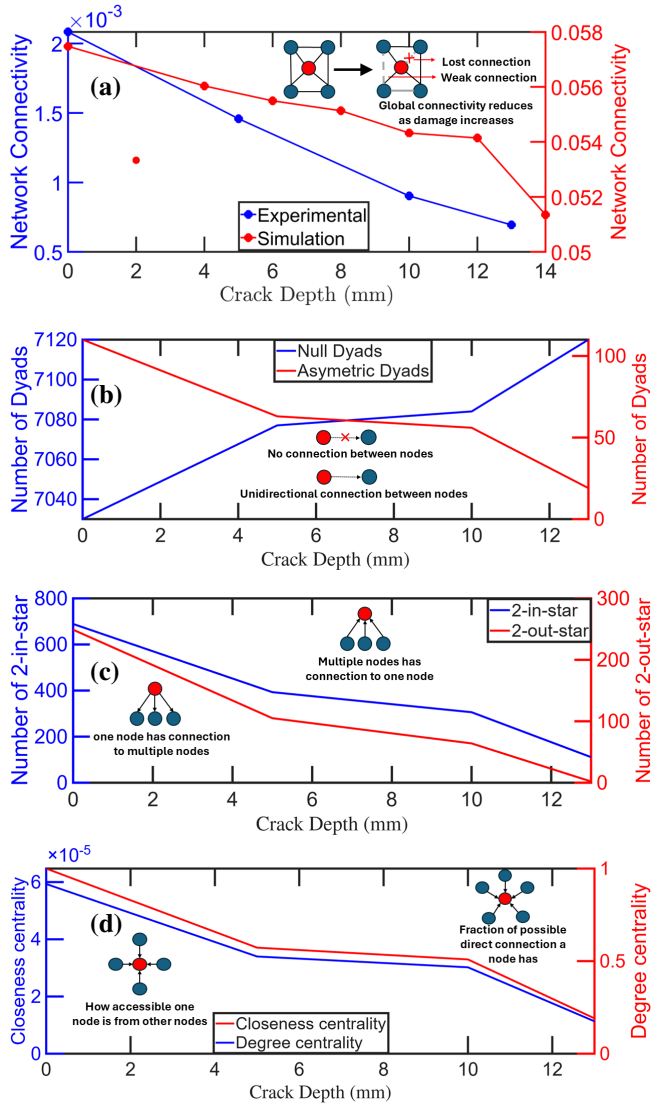
\begin{figure}[t]
    \centering
\vspace{5mm}
    \begin{picture}(240,400)

        \put(0,300){\includegraphics[width=0.95\columnwidth]{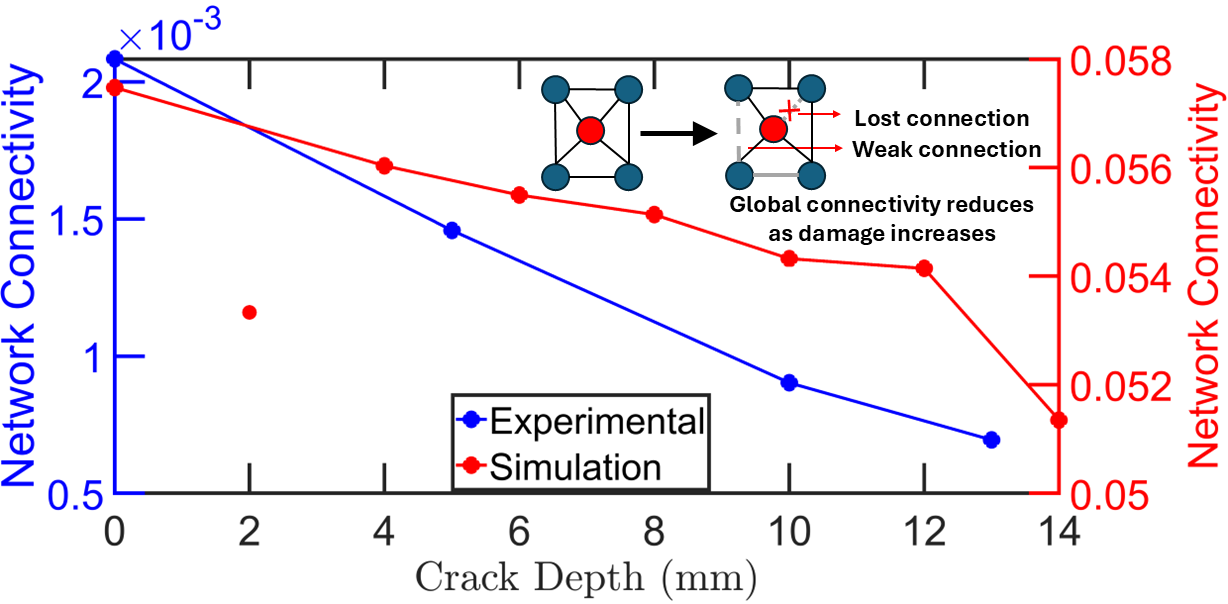}}
        \put(30,380){\normalsize\textbf{(a)}}

        \put(0,200){\includegraphics[width=0.98\columnwidth]{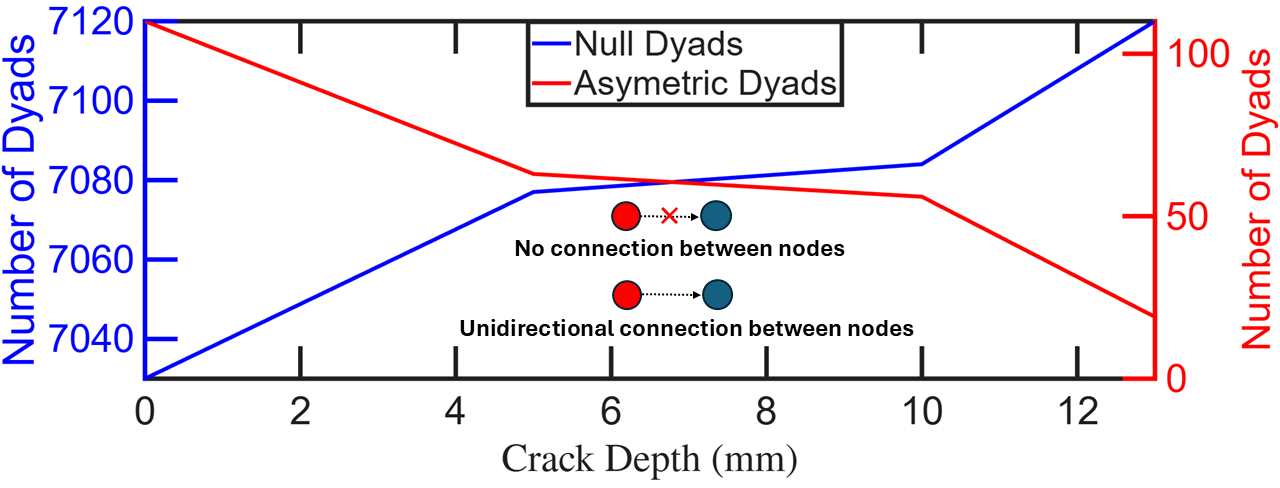}}
        \put(30,264){\normalsize\textbf{(b)}}

        \put(0,100){\includegraphics[width=0.98\columnwidth]{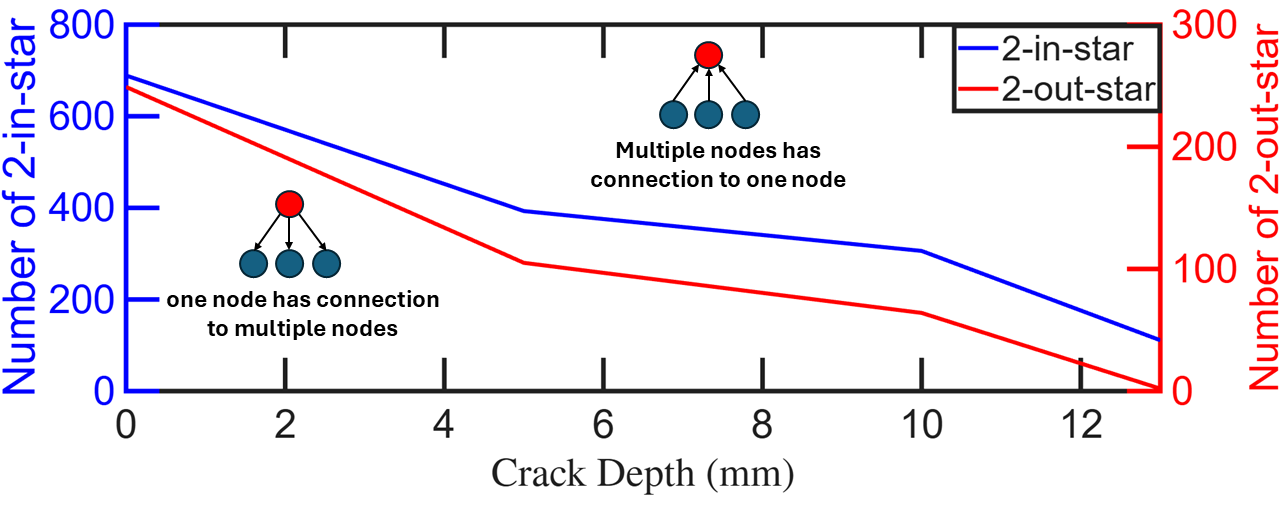}}
        \put(30,160){\normalsize\textbf{(c)}}

        \put(0,0){\includegraphics[width=0.98\columnwidth]{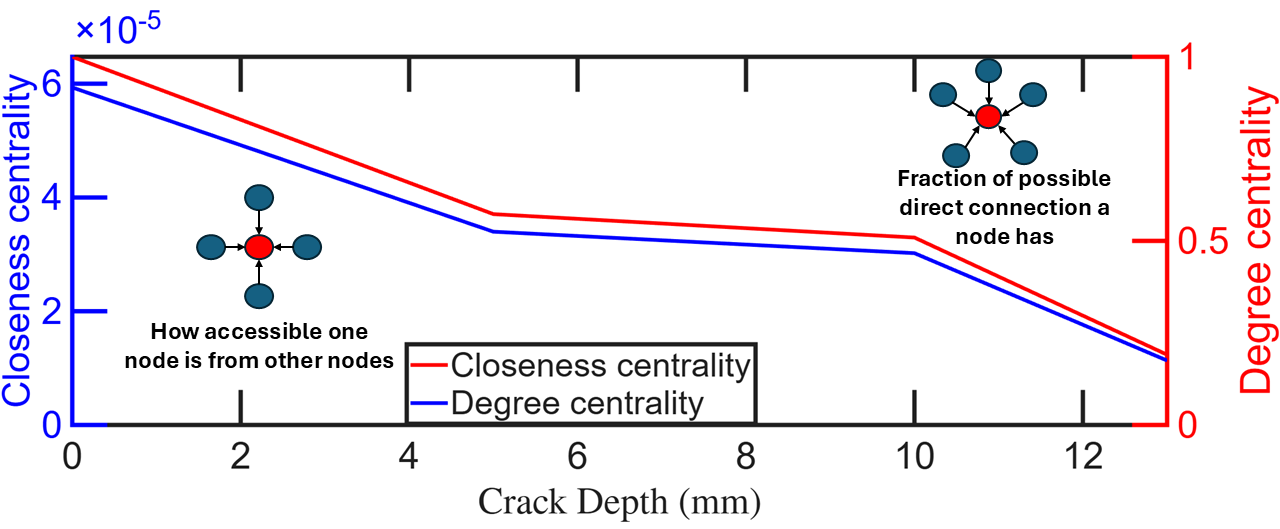}}
        \put(30,65){\normalsize\textbf{(d)}}

    \end{picture}

    \caption{Evolution of the graph theoretic measures with increasing crack depth: (a) network connectivity; (b) dyadic model; (c) higher-order model; and (d) centrality measures.}

    \label{fig:network_parameters}
    \vspace{-5mm}
\end{figure}

To convert the weighted adjacency matrices $\mathbf{G}^{(c)}$ into binary adjacency matrices $\mathbf{B}^{(c)}$, an initial threshold sensitivity analysis was performed over the ranges $\tau \in [0.0165,,0.0200]$ for the experimental cases and $\tau \in [0.07105,,0.07119]$ for the simulation cases. Since the resulting network statistics exhibited consistent trends across these ranges, fixed thresholds of $\tau=0.02$ and $\tau=0.0711$ were selected for the experimental and simulation cases, respectively. Fig.~\ref{fig:network_parameters} illustrates the resulting topological changes in the network as the structural condition evolves with increasing damage.

Fig.~\ref{fig:network_parameters}(a) shows a consistent decrease in network connectivity with increasing crack depth for both the experimental and simulation cases. The highest connectivity is observed in the healthy condition, with a value of $2\times10^{-3}$ and $0.058$ for the experimental and simulation cases, respectively. As damage increases, the connectivity decreases to its lowest values of $0.6\times10^{-3}$ experimentally and $0.054$ in the simulation. Although the experimental and simulation networks have different topological configurations, both exhibit a consistent reduction in connectivity.

While connectivity provides a global description of the network topology, the dyadic and higher-order statistics reveal corresponding changes in local connection patterns. Fig.~\ref{fig:network_parameters}(b) shows that the null dyad count increases from its minimum value of $7030$ in the healthy condition to a maximum of $7120$ at $13,\mathrm{mm}$ crack depth, while the asymmetric dyad count decreases from a maximum of $110$ to a minimum of $19$. A similar decreasing trend is observed in the higher-order structures in Fig.~\ref{fig:network_parameters}(c), where the 2-out-star count decreases from $249$ to $2$ and the 2-in-star count decreases from $689$ to $11$. These reductions indicate that fewer nodes retain the ability to serve as common sources or common receivers for multiple connections. Overall, the decrease in asymmetric dyads and two-star structures, together with the increase in null dyads, indicates a progressive loss of connections and the development of a sparser network with increasing damage severity.

The centrality measures in Fig.~\ref{fig:network_parameters}(d) exhibit a similar damage-dependent trend. As the crack depth increases, degree centrality decreases from $0.9167$ to $0.1750$, while closeness centrality decreases from $6.47\times10^{-5}$ to $1.23\times10^{-5}$. The reduction in degree centrality indicates that individual nodes maintain fewer direct node-to-node connections, whereas the decrease in closeness centrality reflects reduced accessibility through the remaining network paths. Physically, these changes indicate that the DMD-derived spatial relationships become progressively less interconnected as damage increases. The simultaneous reductions in degree and closeness centrality therefore reflect a loss of both direct connectivity and network-wide accessibility in the identified beam dynamics, consistent with the decreasing connectivity trend observed in Fig.~\ref{fig:network_parameters}(a). Overall, with the increase in crack depth, the network evolved from a well-connected structure toward a more fragmented, less accessible, and less locally organized state. These topological changes may be associated with the local stiffness reduction introduced by the crack and the resulting modification of the beam dynamics.

%\subsection{DMD-Based Parameters for Damage Detection}

\begin{figure}[t]
    \centering
\vspace{5mm}
    \begin{overpic}[width=\columnwidth]
        {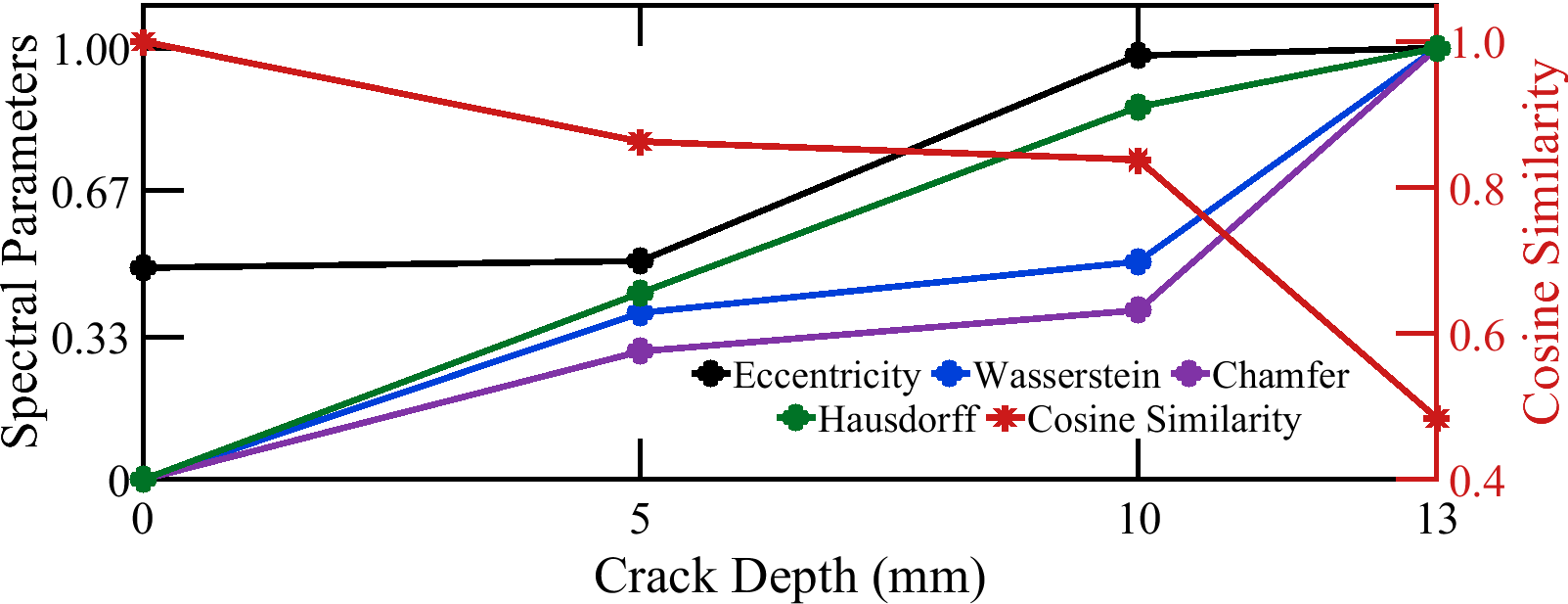}

    \end{overpic}
    
    \caption{Evolution of the DMD-Based measures for damage diagnosis.}
\vspace{-5mm}
    \label{fig:dmd_damage_parameters}
\end{figure}

Fig.~\ref{fig:dmd_damage_parameters} illustrates the evolution of the DMD-based spectral and modal measures as the structural condition changes with increasing crack depth. Eigenspectrum eccentricity is lowest at $0.429$ for the healthy case and rises monotonically to $0.875$ at $13\,\mathrm{mm}$. The increase in eccentricity with crack depth reflects corresponding variations in the decay rates and oscillation frequencies represented by the retained DMD poles. The Wasserstein, Chamfer, and Hausdorff distances also increase with crack depth, starting from zero for the healthy reference and reaching $3.48$, $3.01$, and $5.64$, respectively, at $13\,\mathrm{mm}$ crack. While the Wasserstein distance reflects the collective spectral change and the Chamfer distance represents the average nearest-neighbor distance, the Hausdorff distance captures the largest individual pole deviation and therefore highlights the most pronounced spectral change.

The matched-mode cosine similarity decreases with increasing crack depth, from its highest value of $1.0$ for the healthy reference to $0.484$ at $13\,\mathrm{mm}$ crack. Because structural damage can modify the spatial characteristics of vibration modes through changes in structural properties, the decreasing similarity indicates increasing deviation of the identified DMD modes from the healthy condition. The relatively small reduction at $5$ and $10\,\mathrm{mm}$ suggests that the dominant spatial modal characteristics remain comparatively similar for the shallower cracks, whereas the pronounced decrease at $13\,\mathrm{mm}$ crack indicates a substantially larger modal change. Together, these spectral and modal parameters provide complementary evidence of damage-induced changes in the temporal and spatial characteristics of the identified dynamics, consistent with the topological changes observed in the DMD-derived graph.

\section{Conclusions} 
\label{sec:conclusion}

This study developed a graph theoretic framework for vision-based structural damage identification using delay-embedded DMD. The graph theoretic measures showed a transition toward a less connected network as damage increased, characterized by decreasing connectivity, centrality, and two-star statistics, and an increasing null dyads. The numerical analysis further validated the decreasing trend in graph connectivity, which was observed experimentally. The DMD-based measures provided complementary support to the graph theoretic assessment, exhibiting damage-dependent changes in both the eigenspectrum and modes with increasing crack depth. These results demonstrate that the graph theoretic measures provide reliable indicators for identifying structural damage from vision-based measurements.

\section*{Acknowledgments}

We acknowledge the startup funding support provided by the Mechanical Engineering department at South Dakota State University.

\bibliographystyle{unsrt}     
\bibliography{ACC_references}  

\end{document}